\documentclass[twocolumn,superscriptaddress,amssymb,aps,pra,longbibliography]{revtex4-1}
\usepackage{amsmath,amsfonts,amssymb,amsthm,graphics,graphicx,epsfig,bbm}
\usepackage[colorlinks=true,citecolor=blue,linkcolor=blue,urlcolor=blue]{hyperref}
\usepackage[usenames]{color}

\usepackage{dsfont}
\usepackage{mathrsfs}
\usepackage{amsmath}
\usepackage{subfigure}
\usepackage{braket}
\usepackage{xcolor}
\usepackage{mathtools}
\usepackage{diagbox}

\def\beq{\begin{equation}}
\def\eeq{\end{equation}}
\def\beqa{\begin{eqnarray}}
\def\eeqa{\end{eqnarray}}
\def\bal#1\eal{\begin{align}#1\end{align}}
\def\bfig{\begin{figure}}
\def\efig{\end{figure}}

\newcommand{\eq}[1]{Eq.(#1)}

\newcommand{\fig}[1]{Fig. #1}

\begin{document}
\title{Dissipation-induced nonreciprocal magnon blockade in a magnon-based hybrid system}
\author{Yimin Wang}
\affiliation{Communications Engineering College, Army Engineering University, Nanjing, Jiangsu 210007, China}

\author{Wei Xiong}
\affiliation{Department of Physics, Wenzhou University, Zhejiang 325035, China}

\author{Zhiyong Xu}
\affiliation{Communications Engineering College, Army Engineering University, Nanjing, Jiangsu 210007, China}

\author{Guo-Qiang Zhang}
\email{zhangguoqiang3@zju.edu.cn}
\affiliation{Interdisciplinary Center of Quantum Information, State Key Laboratory of Modern Optical Instrumentation, and Zhejiang Province Key Laboratory of Quantum Technology and Device, Department of Physics, Zhejiang University, Hangzhou 310027, China}

\author{J. Q. You}
\email{jqyou@zju.edu.cn}
\affiliation{Interdisciplinary Center of Quantum Information, State Key Laboratory of Modern Optical Instrumentation, and Zhejiang Province Key Laboratory of Quantum Technology and Device, Department of Physics, Zhejiang University, Hangzhou 310027, China}

\begin{abstract}
We propose an experimentally realizable nonreciprocal magnonic device at the single-magnon level by exploiting magnon blockade in a magnon-based hybrid system. The coherent qubit-magnon coupling, mediated by virtual photons in a microwave cavity, leads to the energy-level anharmonicity of the composite modes. In contrast, the corresponding dissipative counterpart, induced by traveling microwaves in a waveguide, yields inhomogeneous broadenings of the energy levels. As a result, the cooperative effects of these two kinds of interactions give rise to the emergence of the direction-dependent magnon blockade. We show that this can be demonstrated by studying the equal-time second-order correlation function of the magnon mode. Our study opens an avenue to engineer nonreciprocal magnonic devices in the quantum regime involving only a small number of magnons.
\end{abstract}
\date{\today}

\maketitle
\section{Introduction}\label{introduction}

Owing to great application prospects in quantum technologies, a new class of hybrid quantum systems based on magnons, i.e., the collective spin excitations in ferro- and ferrimagnetic crystals such as yttrium iron garnet (YIG), have blossomed in the past few years~\cite{Rameshti2021,PengYan-2021,Lachance-Quirion2019}, where magnons are coherently coupled to microwave potons~\cite{Huebl2013,Cao2015,yuan2017apl,Tabuchi2014,Zhang2014,Goryachev2014,Hu-PRL-15,Zhang-npjQI-15,Wang18}, phonons~\cite{Zhang2016,Gao2017,Lu2021,Potts2021}, optical photons~\cite{Haigh2016,Haigh2015}, and superconducting qubits~\cite{tabuchi2015s,Quirion20}. Benefiting from the advantages of intrinsically good tunability as well as long coherence time, many novel phenomena have been explored in these magnon-based hybrid systems~\cite{Nair21,Shim20,Zhang-China-19,Bi21,Yu19,You19,Yuan20,huai2019pra,Zhang2021-PRB,Grigoryan18,Li18}. For example, the nonreciprocal microwave transmission (i.e., allowing the flow of signal to propagate from one side but not the other side) in the linear regime has been achieved with the assistance of the dissipative coupling between magnons and microwave photons~\cite{wang2019prl,Qian20}. Nonreciprocal devices are indispensable in a wide range of practical applications, such as optical isolators~\cite{Fiederling99,Shintaku98}, metamaterials~\cite{Mahmoud15,Hamann18}, and circulators~\cite{Scheucher16}.

In addition, the magnon blockade, in analogy to photon blockade~\cite{Imamoglu1997,Liew10} and phonon blockade~\cite{Liu10,Wang2016}, was theoretically proposed in Refs.~\cite{liu2019prb,Xie20}, where magnons in a YIG sample are strongly coupled to a superconducting qubit. In such a qubit-magnon hybrid system, the anharmonic energy-level structure of the system prevents the resonant injection of more than one magnon into the magnon mode (i.e., the magnon blockade occurs)~\cite{tabuchi2015s,Quirion20}. The magnon blockade is a pure quantum effect, which provides the possibility to develop novel magnonic devices at the single-magnon level, such as single magnon sources. Very recently, Refs.~\cite{Huang18} and \cite{Yao2021} have proposed to engineer the nonreciprocal photon blockade in a spinning Kerr cavity and the nonreciprocal phonon blockade in a composite spin-phononic system, respectively. In chiral quantum technologies, quantum nonreciprocal devices are crucial elements and have received extensive attention~\cite{Yang2019,Jiao2020,Dong2021,Xu2020,Xu2021}. However, up to now, the nonreciprocal magnon blockade has not yet been investigated.

In this work, we propose to engineer nonreciprocal magnonic devices at the single-magnon level. In our approach, the magnon mode in a YIG sphere is both coherently and dissipatively coupled to a superconducting qubit. Due to the interference between coherent and dissipative qubit-magnon couplings, the non-classical statistics of magnons depends on the direction of the drive microwave field of the coupled hybrid system. In such a system, for the right-propagating (or left-propagating) drive field, the magnon statistics can be sub-Poissonian, corresponding to the occurrence of magnon blockade. However, when reversing the propagation direction of the drive field, the magnon statistics becomes super-Poissonian, indicating the disappearance of the magnon blockade. This work opens a route to engineer nonreciprocal magnonic devices in the quantum regime, which may have important application aspects in the versatile magnon-based quantum information processing platforms~\cite{Rameshti2021,PengYan-2021,Lachance-Quirion2019}.

\begin{figure}[b!]
\begin{center}
\includegraphics[scale=0.44]{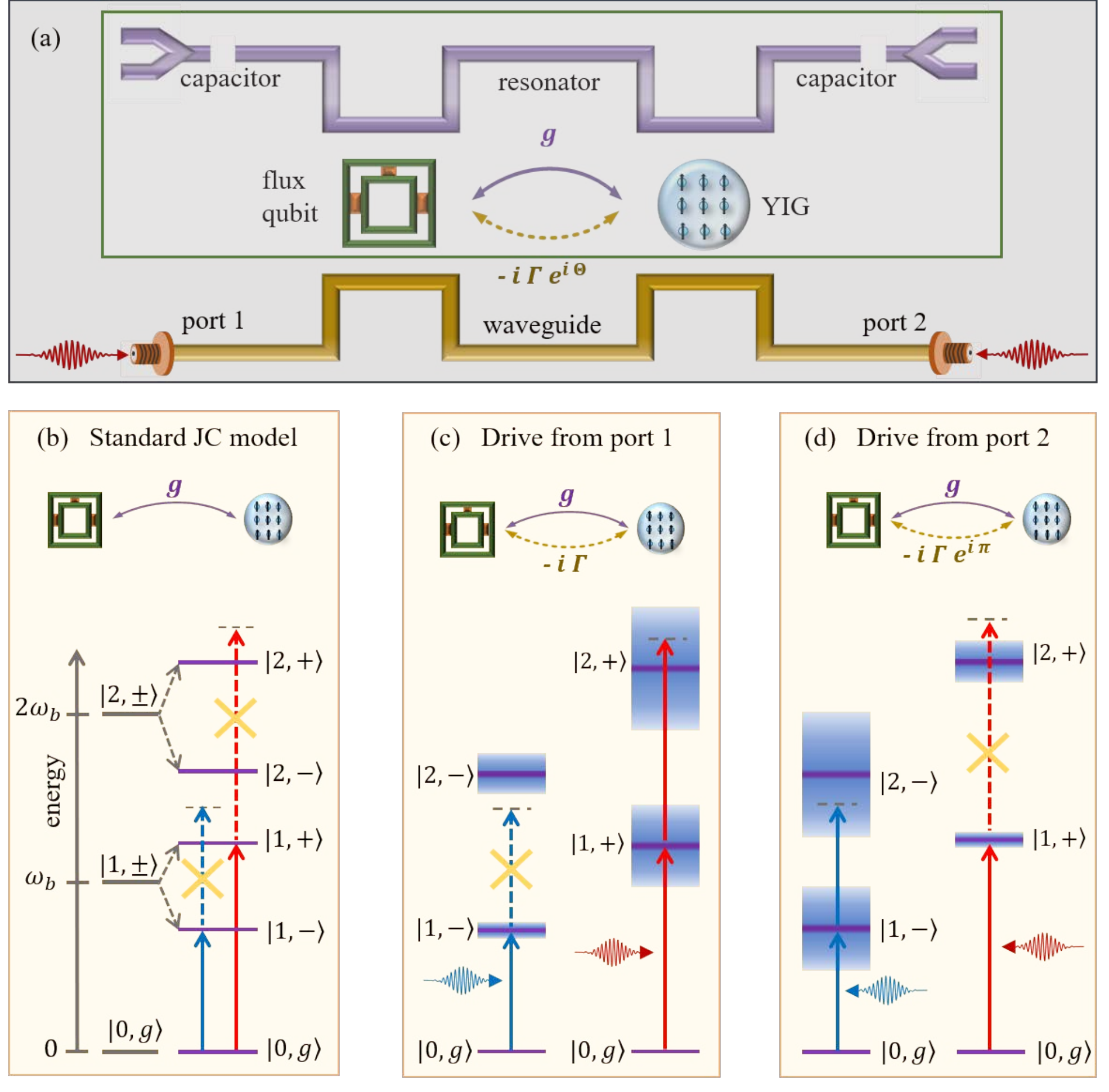}
\end{center}
\caption{(a) Schematic of the proposed hybrid quantum system for nonreciprocal magnon blockade. A transmission-line resonator, which supports standing cavity modes, is employed to mediate the coherent coupling between magnons in a ferrimagnetic YIG sphere and the superconducting flux qubit (see the green box). In addition, both the mangon mode and the flux qubit are also coupled to the traveling waves in an open waveguide, resulting in the qubit-magnon dissipative coupling with strength $\Gamma$. The microwave signals are loaded via port 1 and port 2 for drivings. (b) Energy levels of the qubit-magnon hybrid system for illustrating the magnon blockade in the standard Jaynes-Cummings model, where the coupling between the magnon mode and the qubit only involves the coherent part, without the dissipative contribution to the coupling. (c) and (d) Schematic demonstration of the nonreciprocal magnon blockade with a driving loaded via either (c) port 1 or (d) port 2 in the qubit-magnon system, where both the coherent and dissipative couplings are included.}
\label{fig1}
\end{figure}

\section{The model}

As schematically depicted in the green box in \fig{\ref{fig1}}(a), the hybrid quantum system consists of a YIG sphere, a superconducting flux qubit and a coplanar transmission-line resonator (called the cavity here), where both the qubit and the considered magnon mode (i.e., the Kittel mode with uniform spin precession) in the YIG sphere are strongly coupled to the microwave cavity mode. When the cavity has a large frequency detuning from both the qubit and the magnon mode, the effective strong interaction between the qubit and the magnon mode can be realized via exchanging virtual microwave photons in the cavity~\cite{tabuchi2015s,Quirion20}. By eliminating the degrees of freedom of the cavity mode via the Fr\"{o}hlich-Nakajima transformation~\cite{Frohlich50,Nakajima55}, we obtain the effective Hamiltonian of the qubit-magnon hybrid quantum system (set $\hbar$ = 1; see Appendix \ref{Appendix-A} for details)
\begin{eqnarray}\label{JC}
H= \omega_q\sigma_+ \sigma_-+\omega_b b^{\dag}b+\lambda\left(b^{\dag}\sigma^{-}+b\sigma^{+}\right),	
\end{eqnarray}
where $\sigma_+ = \ket{e}\bra{g}$ ($\sigma_- = \ket{g}\bra{e}$) is the raising (lowering) operator of the qubit with ground state $\ket{g}$ and excited state $\ket{e}$, and $b^\dag$ ($b$) is the creation (annihilation) operator of the magnon mode. The bare frequency $\omega^{(0)}_q$ ($\omega^{(0)}_b$) of the qubit (magnon mode) is slightly modified to $\omega_q = \omega^{(0)}_q +\lambda_q^2/\delta_q$ ($\omega_b=\omega_b^{(0)} +{\lambda_b^2}/\delta_b$), and $\lambda={\lambda_q\lambda_b}({1}/{2\delta_q}+{1}/{2\delta_b})$ is the effective coherent coupling strength between the qubit and the magnon mode. Here $\delta_{q(b)}=\omega^{(0)}_{q(b)}-\omega_c$ is the frequency difference between the qubit (magnon mode) and the cavity field with frequency $\omega_c$, whereas $\lambda_q$ ($\lambda_b$) stands for the corresponding coupling strength.

In the Hermitian case, i.e., without the dissipation in the qubit-magnon system, the eigenstates of the qubit-magnon system include the ground state $\ket{0,g}\equiv \ket{0}\ket{g}$, and the excited eigenstates $\ket{n,\pm}$:
\begin{eqnarray}\label{eigenstates}
\ket{n,-}&=&\cos\theta_n\ket{n}\ket{g}-\sin\theta_n\ket{n-1}\ket{e},\nonumber\\
\ket{n,+}&=&\cos\theta_n\ket{n-1}\ket{e}+\sin\theta_n\ket{n}\ket{g}.
\end{eqnarray}
The corresponding eigenenergies are $\omega_{0,g}=0$, and
\begin{eqnarray}\label{eigenstates}
\omega_{n,\pm}=n\omega_b+\frac{1}{2}\delta \pm \frac{1}{2}\sqrt{\delta^2+4n\lambda^2},
\end{eqnarray}
where $\ket{n}$ $(n=0,1,2,...)$ are the Fock states of the magnon mode, $\delta=\omega_q-\omega_b$ is the frequency detuning of the qubit relative to the magnon mode, and $\theta_n=\tan^{-1}(2\lambda\sqrt{n}/\delta)/2$. When the qubit is near-resonantly and strongly coupled to the magnon mode (i.e., $\omega_q \approx\omega_b$ and $\lambda \gg \{\gamma_{\rm in},\kappa_{\rm in}\}_{\rm max}$, where $\gamma_{\rm in}$ is the intrinsic decay rate of the qubit and $\kappa_{\rm in}$ is the intrinsic decay rate of the magnon mode), the strong anharmonicity of the energy spectrum yields the magnon blockade in the weak driving limit~\cite{liu2019prb,Xie20}, as schematically shown in Fig.~\ref{fig1}(b). The transitions between the quantum states $|0,g\rangle$ and $|1,\pm \rangle$ are permitted as long as the drive field has the resonant frequencies of $\omega_d \approx \omega_b \pm \lambda$, respectively. Meanwhile, the subsequent transitions $|1,\pm\rangle \rightarrow |2,\pm \rangle$ are inhibited, due to the large frequency detuning $|(\omega_{2,\pm}-\omega_{1,\pm})-\omega_d| \approx (2-\sqrt{2})\lambda$~($>\{\gamma_{\rm in},\kappa_{\rm in}\}_{\rm max}$). In other words, regardless of the driving direction, once a magnon is excited in the system, it prevents the second magnon with the same frequency from entering the qubit-magnon system.

To engineer the nonreciprocal magnon blockade, we specially extend the cavity-mediated qubit-magnon system to include an open waveguide (i.e., without the effective mirrors to form a cavity), as schematically demonstrated in \fig{\ref{fig1}}~(a). In the extended qubit-magnon system, the waveguide acts as a common reservoir, and both the qubit and the magnon mode individually interact with the traveling microwaves in the waveguide~\cite{metelmann2015prx,wang2019prl}. To excite the magnons in the YIG sphere, a weak microwave field with frequency $\omega_d$ is loaded into the qubit-magnon system through either port 1 or port 2, located at the two ends of the waveguide. Mediated by the waveguide, the loaded external microwave field drives both the qubit and the magnon mode in the following form~\cite{wang2019prl,Qian20}:
\begin{eqnarray}\label{drive}
H_d&=&\xi_q( \sigma_-e^{i\omega_d t} +\sigma_+ e^{-i\omega_d t}) \nonumber\\
         & &+\xi_b[be^{i(\omega_d t+\phi)} +b^\dagger e^{-i(\omega_d t+\phi)}],
\end{eqnarray}
where $\xi_q=\sqrt{\gamma_{\rm ex}}\,\xi$ ($\xi_b=\sqrt{\kappa_{\rm ex}}\,\xi$) is the Rabi frequency related to the qubit (magnon mode), $\xi$ represents the drive strength of the microwave field, $\gamma_{\rm ex}$ ($\kappa_{\rm ex}$) characterizes the coupling strength between the qubit (magnon mode) and the waveguide, and the phase $\phi = k L$ (with $L$ being the separation distance between the qubit and the YIG sphere and $k$ being the wave number) indicates the relative phase delay of traveling photons from the qubit to the magnon mode. Hereafter, we choose $\phi =0$ in our work.

When including the external drive field and the dissipations of the system, the dynamics of the total qubit-magnon system can be described by a Lindblad master equation~\cite{metelmann2015prx,breuer2007}:
\begin{eqnarray} \label{eq_me}
\dot{\rho} =i \left[\rho, H+H_d\right]+\gamma_{\rm in}{\mathcal L}[\sigma_-]\,\rho +\kappa_{\rm in}{\mathcal L}[b]\,\rho + \tau {\mathcal L}[o] \,\rho,~~~~~
\end{eqnarray}
where the standard dissipative Lindblad superoperator has the form
\begin{eqnarray}\label{eq_linL}
{\mathcal L}[f] \,\rho=2f \rho f^{\dagger}-f^{\dagger} f \rho-\rho f^\dagger f,
\end{eqnarray}
with $f=\{\sigma_-,\,b,\,o\}$. On the right side of Eq.~(\ref{eq_me}), the second and third terms describe the intrinsic dissipations of the qubit and the magnon mode, respectively, and the fourth term represents the cooperative dissipation of the qubit and the magnon mode into the waveguide with decay rate $\tau$. The jump operator $o$ is generally expressed as the superposition of both qubit's and magnon's operators, and more precisely, it depends on which port the drive field is loaded into~\cite{wang2019prl,Qian20}. In the case with the drive field loaded into the qubit-magnon system via port 1, the jump operator can be defined as $o=\mu b + \nu \sigma_-$~\cite{metelmann2015prx}, with the coefficients $\mu$ and $\nu$ satisfying $\tau \cdot \nu^{2}=\gamma_{\rm ex}$ and $\tau \cdot \mu^{2}=\kappa_{\rm ex}$, which characterize the individual couplings of the qubit and magnon to the traveling waves in the open waveguide. However, when the drive field is loaded via port 2, the jump operator becomes $o=-\mu b  + \nu \sigma_-$, since the propagation direction of the traveling waves in the waveguide is reversed~\cite{wang2019prl,Qian20}. For convenience, we introduce a phase $\Theta$ to indicate the difference in the drive direction, i.e.,
\begin{eqnarray}\label{eq_opo}
o = e^{i\Theta}\mu b + \nu \sigma_- ,
\end{eqnarray}
where $\Theta=0$ and $\Theta=\pi$ correspond to the microwave drive loaded via port 1 and port 2, respectively.

Neglecting the quantum jump terms $2\gamma_{\rm in}\sigma_{-}\rho\sigma_{+}$, $2\kappa_{\rm in} b\rho b^\dag$ and $2\tau o\rho o^\dag$, we can write the master equation in Eq.~({\ref{eq_me}}) as the Liouvillian equation~\cite{Minganti2019,Zhang2021-PRX}: $\dot{\rho} \approx i[\rho(H_{\rm eff}+H_d)^\dag-(H_{\rm eff}+H_d)\rho]$, where
\begin{eqnarray}\label{eq_ht1eff}
H_{\rm eff} &=& (\omega_q - i \gamma)\sigma_+ \sigma_-+(\omega_b - i \kappa) b^{\dagger}b\nonumber \\
              & &+ (\lambda-i\Gamma e^{i \Theta})(\sigma_{+}b +b^{\dagger} \sigma_{-})
\end{eqnarray}
is the effective non-Hermitian Hamiltonian of the qubit-magnon system, which approximatively determines the non-classical statistics of magnons in the proposed qubit-magnon system. Here, $\gamma = \gamma_{\rm in}+\gamma_{\rm ex}$ ($\kappa =\kappa_{\rm in}+\kappa_{\rm ex}$) represents the total decay rate of the qubit (magnon mode), and $\Gamma=\sqrt{\kappa_{\rm ex}\gamma_{\rm ex}}$ denotes the effective dissipative coupling between the qubit and the magnon mode. For the cases with the drive field injected via port 1 and port 2, respectively, the phases of the dissipative coupling $\Gamma$ relative to the coherent coupling $\lambda$ differs by $\pi$, i.e., $H_{\rm eff}$ is nonreciprocal. This drive direction-dependent nonreciprocity substantially leads to the nonreciprocal magnon blockade, which can be intuitively explained using the complex energy spectrum of the effective Hamiltonian $H_{\rm eff}$ in Eq.~(\ref{eq_ht1eff}) (cf. Sec.~\ref{NMB}).

\begin{figure}[b]
\begin{center}
\includegraphics[scale=0.7]{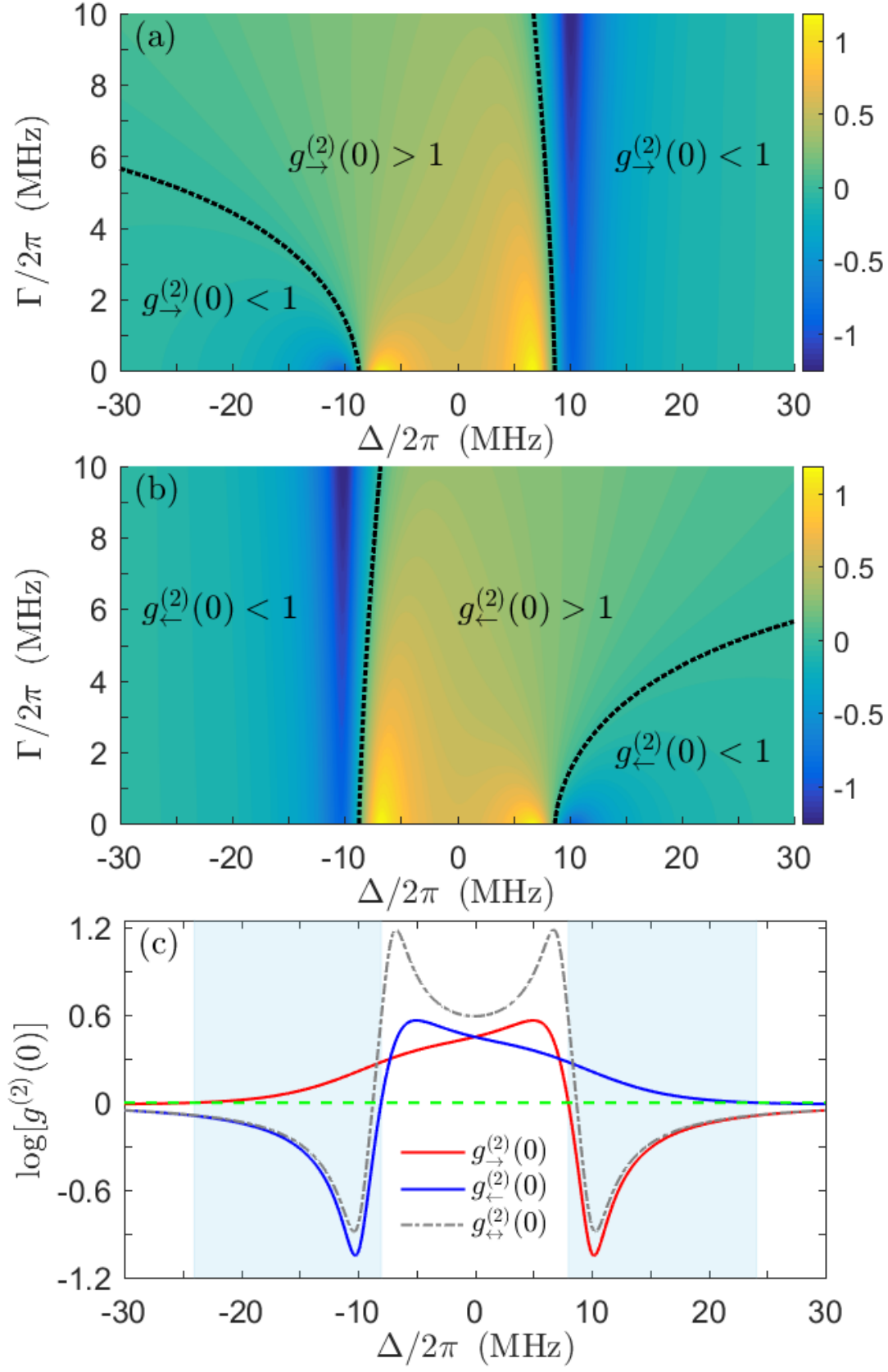}
\end{center}
\caption{(a) For the right-propagating drive field injected via port 1 with $\Theta=0$ and (b) the left-propagating drive field imported via port 2 with $\Theta=\pi$, the steady-state logarithmic equal-time second-order correlation functions to the base of $10$, i.e., $\log_{10}[g_{\rightarrow}^{(2)}(0)]$ and $\log_{10}[g_{\leftarrow}^{(2)}(0)]$, are plotted versus the detuning $\Delta/2\pi$ and the dissipative coupling strength $\Gamma/2\pi$. The black dotted curves are plotted for $g^{(2)} (0)=1$, which are the boundaries between sub-Poissonian [$g^{(2)}(0) < 1$] and super-Poissonian [$g^{(2)}(0) >1$] magnon statistics. Other parameters are chosen as $\omega_b/2\pi = \omega_q/2\pi = 5$~GHz, $\lambda/2\pi=10$~MHz, $\gamma_{\rm in}/2\pi = \kappa_{\rm in}/2\pi = 1$~MHz, $\mu=\nu=1$, and $\tau=\Gamma$. (c) Cross section views of the second-order correlation function $\log_{10}[g^{(2)}(0)]$ obtained by slicing the entire range of the detuning $\Delta/2\pi$ at $\Gamma/2\pi=5$~MHz for different input directions, i.e., $g_{\rightarrow}^{(2)}(0)$ marked in red and $g_{\leftarrow}^{(2)}(0)$ marked in blue. The gray dash-dotted curve shows the direction-independent correlation functions $g_{\leftrightarrow}^{(2)}(0)$ with $\tau=0$ for comparison, and the green dashed line represents the reference level of $g^{(2)}(0)=1$.}
\label{fig2}
\end{figure}

\section{Nonreciprocal magnon blockade based on the coherent and dissipative qubit-magnon couplings}\label{NMB}

\subsection{Nonreciprocal magnon blockade}

To quantitatively characterize the non-classical statistics of magnons, we introduce the equal-time second-order correlation function~\cite{Scully97},
\begin{eqnarray} \label{eq_gn}
g^{(2)}(0) \equiv \frac{\langle b^\dag b^\dag bb\rangle}{\langle b^\dag b\rangle^2}
= \frac{{\rm Tr} [\rho_{\rm ss}b^\dagger b^\dag bb ]}{{\rm Tr} [\rho_{\rm ss}b^\dagger b]^2},
\end{eqnarray}
where $\rho_{\rm ss}$ is the density matrix of the qubit-magnon system in the steady state. The condition of $g^{(2)}(0) <1$ or $g^{(2)}(0) >1$ indicates the appearance of sub-Poissonian or super-Poissonian magnon statistics.
When $g^{(2)}(0) <1$, magnon blockade occurs~\cite{liu2019prb,Xie20}, which is a pure quantum phenomenon. The limit of $g^{(2)}(0) \rightarrow 0$ predicts the perfect magnon blockade, i.e., only one magnon can be excited in the qubit-magnon system.

In what follows, we focus on the typical second-order correlation functions $g_{\rightarrow}^{(2)}(0)$ and $g_{\leftarrow}^{(2)}(0)$, where the arrows $\rightarrow$ and $\leftarrow$ in the subscripts denote the right- and left-propagating drive field injected via the port 1 and the port 2, respectively, corresponding to $\Theta=0$ and $\Theta=\pi$. Without the loss of generality, identical frequencies are assumed in numerical simulations for the qubit and the magnon mode, i.e., $\omega_q = \omega_b$. In Figs.~{\ref{fig2}}(a) and {\ref{fig2}}(b), we demonstrate the steady-state logarithmic (of base 10) $g_{\rightarrow}^{(2)}(0)$ and $g_{\leftarrow}^{(2)}(0)$ versus the dissipative coupling rate $\Gamma$ and the detuning $\Delta=\omega_b-\omega_d$ by numerically calculating Eqs.~(\ref{eq_me}) and (\ref{eq_gn}), where the black dotted curves [$g^{(2)}(0) =1$] denote the boundaries between sub-Poissonian [$g^{(2)}(0) < 1$] and super-Poissonian [$g^{(2)}(0) >1$] magnon statistics. It is obvious from Figs.~{\ref{fig2}}(a) and {\ref{fig2}}(b) that the correlation functions $g_{\rightleftarrows}^{(2)}(0)$ for bidirectional microwave drives own the appealing feature of symmetry about the line $\Delta=0$, i.e., $g_{\rightarrow}^{(2)}(0)|_{\Delta=\Delta_0} = g_{\leftarrow}^{(2)}(0)|_{\Delta=-\Delta_0}$ for any values $\Delta_0$ of the detuning $\Delta$. In the intersection regions of $g_{\rightarrow}^{(2)}(0) >1$ and $g_{\leftarrow}^{(2)}(0) <1$, or conversely, $g_{\rightarrow}^{(2)}(0) <1$ and $g_{\leftarrow}^{(2)}(0) >1$, the extraordinary nonreciprocal magnon blockade emerges [see the light-blue regions in Fig.~{\ref{fig2}}(c)], where super-Poissonian magnons occur by loading the drive field via one port, while sub-Poissonian magnons appear via the other port.

\begin{figure}[bt]
\begin{center}
\includegraphics[scale=0.7]{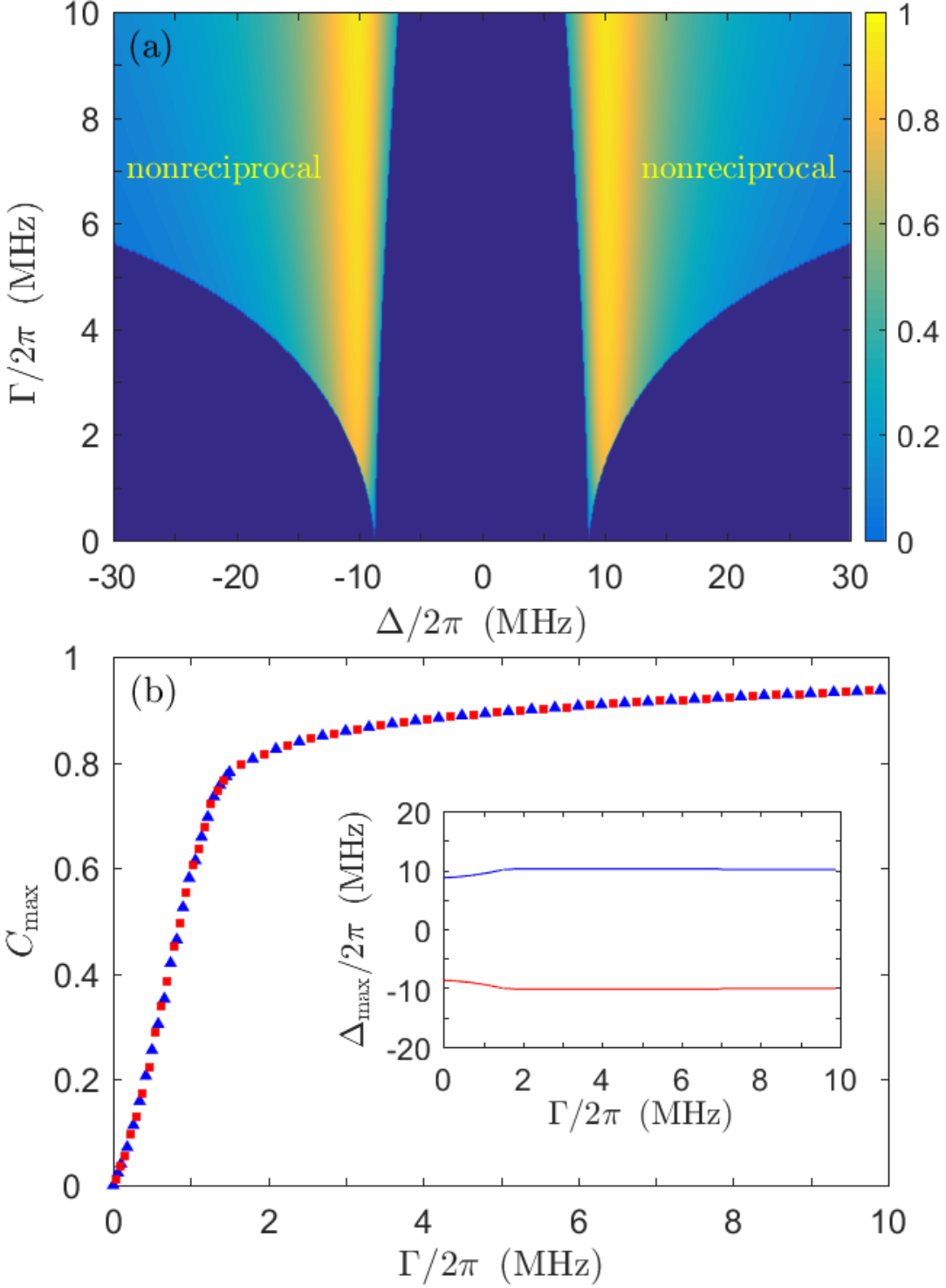}
\end{center}
\caption{(a) Dependence of the bidirectional contrast ratio $C$ on the dissipative coupling strength $\Gamma/2\pi$ and the detuning $\Delta/2\pi$. (b) The maximized contrast ratio $C_{\rm max}$ versus the dissipative coupling strength $\Gamma/2\pi$, with the corresponding detuning $\Delta_{\rm max}/2\pi$ of blue triangles (red squares) indicated by the blue (red) curve in the inset. Other parameters are the same as in Fig.~\ref{fig2}.}
\label{fig3}
\end{figure}

In order to quantitatively describe the nonreciprocal magnon blockade, we introduce the bidirectional contrast ratio $C$ (satisfying $0 \leq C \leq 1$) in the nonreciprocal regimes,
\begin{equation}
C = \left|\frac{g_{\rightarrow}^{(2)}(0)-g_{\leftarrow}^{(2)}(0)}{g_{\rightarrow}^{(2)}(0)+g_{\leftarrow}^{(2)}(0)}\right|,
\end{equation}
where $C=1$ ($C=0$) corresponds to the ideal nonreciprocal magnon blockade (the disappearance of the nonreciprocity). The higher the contrast ratio $C$ is, the stronger the nonreciprocity of the magnon blockade is. From \fig{\ref{fig3}}(a), it is clear that the nonreciprocal magnon blockade occurs in a remarkably broad parameter range and the contrast ratio $C$ is symmetric about the line $\Delta=0$. For a given value of the dissipative qubit-magnon coupling strength $\Gamma$, we define the maximal value of the contrast ratio $C$ as $C_{\rm max}$, where the corresponding detuning $\Delta$ is also denoted as $\Delta_{\rm max}$. In Fig.~{\ref{fig3}}(b), we display the maximal contrast ratio $C_{\rm max}$ versus $\Gamma$, which increases monotonically with $\Gamma$. In the region $\Gamma/2\pi>1.8$~MHz, apparent nonreciprocal magnon blockade scenarios take place with $C_{\rm max}>0.8$.
Regardless of the values of $\Gamma$, the maximum $C_{\rm max}$ is always obtained around $\Delta_{\rm max} \approx \pm \lambda=\pm 10\times 2\pi$~MHz [see the inset of Fig.~{\ref{fig3}}(b)]. For example, when $\Gamma/2\pi = 5$~MHz, the maximal contrast ratio $C_{\rm max}$ reaches $0.895$ at $\Delta/2\pi =- 10.2$~MHz, with $g_{\rightarrow}^{(2)}(0) = 1.615$, and $g_{\leftarrow}^{(2)}(0)= 0.089$.

\begin{figure}[b]
\begin{center}
\includegraphics[scale=0.7]{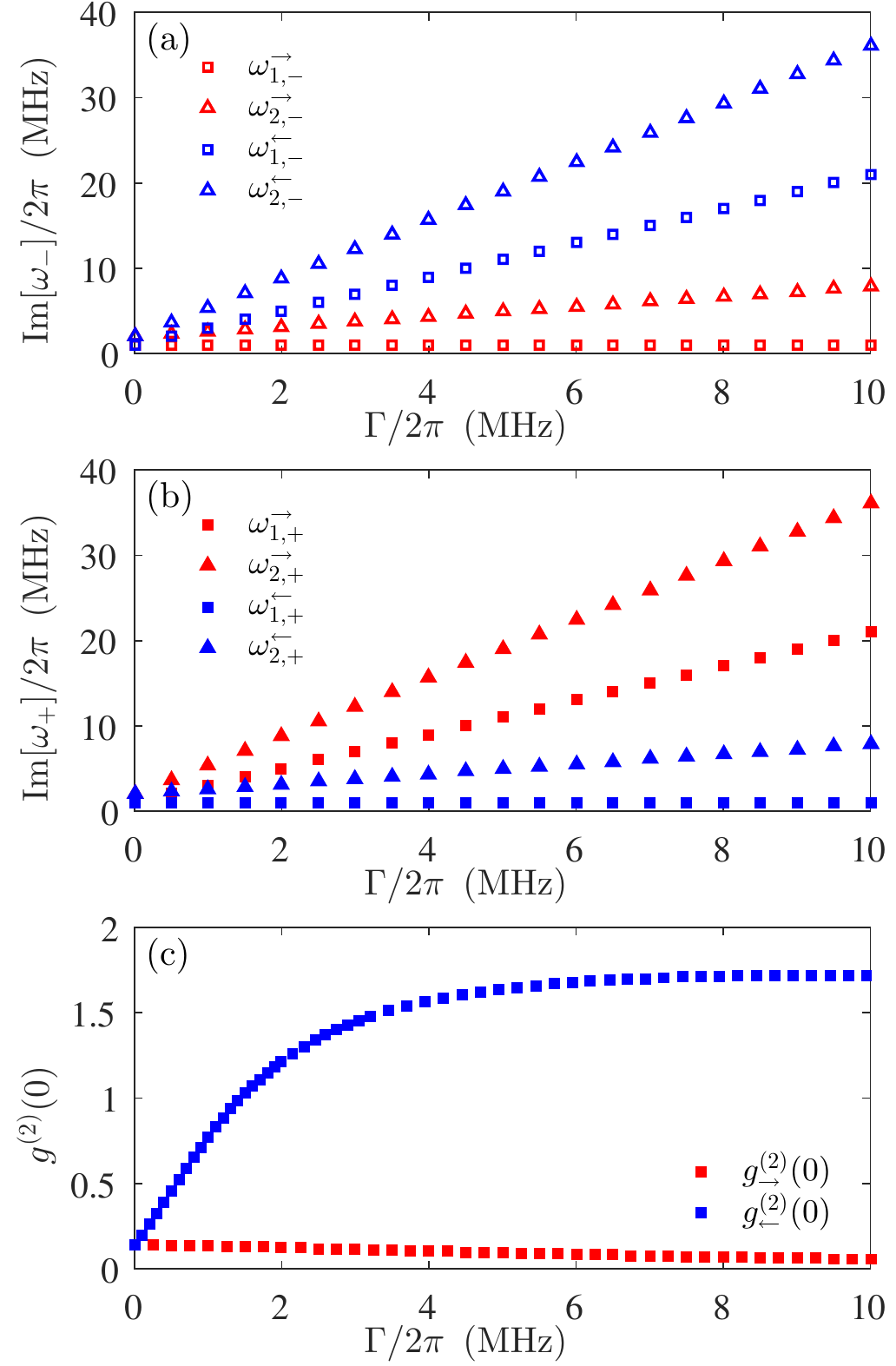}
\end{center}
\caption{Imaginary parts (a) $\rm{Im}[\omega_{n,-}]$ and (b) $\rm{Im}[\omega_{n,+}]$ of the bidirectional eigenvalues $\omega_{n,\pm}$ $(n=1,2)$ versus the dissipative coupling $\Gamma/2\pi$. (c) Bidirectional correlation function $g^{(2)}(0)$ versus $\Gamma/2\pi$ with $\Delta/2\pi =10$~MHz. Other parameters are the same as in Fig.~\ref{fig2}.}
\label{fig4}
\end{figure}

\subsection{Origin of the nonreciprocal magnon blockade}

Below we proceed to investigate the physical mechanism of the nonreciprocal magnon blockade in the qubit-magnon system. Corresponding to Eq.~(\ref{eigenstates}), the eigenenergies of the non-Hermitian hybrid system become
\begin{align}
\omega_{n,\pm}=&n \tilde{\omega}_b+\frac{1}{2}\tilde{\Delta}\pm\frac{1}{2}\sqrt{\tilde{\Delta}^2+4n{(\lambda-i\Gamma e^{i \Theta})}^2},
\label{eqwdress}
\end{align}
with $\tilde{\omega}_b = \omega_b- i \kappa$, $\tilde{\omega}_q = \omega_q - i \gamma$ and $\tilde{\Delta}=\tilde{\omega}_q-\tilde{\omega}_b$, which are obtained by diagonalizing the effective non-Hermitian Hamiltonian $H_{\rm eff}$ in Eq.~(\ref{eq_ht1eff}). In this non-Hermitian qubit-magnon system, the real parts of $\omega_{n,\pm}$, i.e., ${\rm Re}[\omega_{n,\pm}]$, presents the frequencies of the dressed states; the imaginary parts, ${\rm Im}[\omega_{n,\pm}]$, indicate the total dissipations of the energy levels, which result in the extended frequency broadenings of the energy levels. As intuitively demonstrated in Figs.~{\ref{fig1}}(c) and {\ref{fig1}}(d), the anharmonicity of the energy levels of the qubit-magnon system, which is the origin of the magnon blockade~\cite{liu2019prb,Xie20}, is direction-independent, i.e., ${\rm Re}[\omega_{n,\pm}]$ is independent of $\Theta$ [cf. \eq{\ref{eqwdress}}]. That is to say, they are the same for drivings from both directions when $\Theta = \{0,\pi\}$. However, the linewidths ${\rm Im}[\omega_{n,\pm}]$ of these levels become direction-dependent [see Figs.~\ref{fig4}(a) and \ref{fig4}(b)], inherently accounting for the nonreciprocity of the magnon blockade. Moreover, Figs.~\ref{fig4}(a) and \ref{fig4}(b) also reveal the fact that the amounts of inhomogeneous broadening of these levels are monotonically increasing with rising dissipative coupling $\Gamma$. The dissipative coupling due to the traveling waves in the waveguide causes dramatic changes to the second-order correlation functions for drivings from different directions, as presented in Fig.~\ref{fig4}(c).

As schematically shown in Fig.~\ref{fig1}(c), when the drive field comes from the left side (port 1) with frequency $\omega_d = {\rm Re}[\omega_{1,+}]$, the linewidth $|{\rm Im}[\omega_{2,+}]|$ of the state $|2,+\rangle$ becomes comparable to (or larger than) the frequency detuning $|({\rm Re}[\omega_{2,+}]-{\rm Re}[\omega_{1,+}])-\omega_d|$ between the drive field and the transition $|1,+\rangle \rightarrow |2,+\rangle$ (i.e., $|{\rm Im}[\omega_{2,+}]|\geq |({\rm Re}[\omega_{2,+}]-{\rm Re}[\omega_{1,+}])-\omega_d|$). Therefore, in such a case, the state $|2,+ \rangle$ may be excited, and the absorption of the first magnon also favors that of the second or subsequent magnons, giving rise to the super-Poissonian magnon statistics with $g_{\rightarrow}^{(2)}(0)>1$; see the region with $\Delta/2\pi \approx -10$~MHz and $\Gamma/2\pi>2$~MHz in Fig.~\ref{fig2}(a)]. However, sub-Poissonian magnon statistics with $g_{\rightarrow}^{(2)}(0)<1$ remains for the drive field with frequency $\omega_d = {\rm Re}[\omega_{1,-}]$ [see the region with $\Delta/2\pi \approx 10$~MHz in Fig.~\ref{fig2}(a)], since the drive frequency $\omega_d$ is still far detuned from the transition frequency ${\rm Re}[\omega_{2,-}] - {\rm Re}[\omega_{1,-}]$ even in the linewidth-broadening case (i.e., $|{\rm Im}[\omega_{2,-}]|< |({\rm Re}[\omega_{2,-}]-{\rm Re}[\omega_{1,-}])-\omega_d|$), cf. Fig.~\ref{fig1}(c). As a contrast and revealed in Fig.~\ref{fig1}(d), for the drive field imported via port 2, the signatures of magnon statistics for drives with frequencies $\omega_d ={\rm Re}[\omega_{1,\pm}]$ get switched, namely, $g_{\leftarrow}^{(2)}(0)>1$ for $\omega_d ={\rm Re}[\omega_{1,-}]$, while $g_{\leftarrow}^{(2)}(0)<1$ for $\omega_d ={\rm Re}[\omega_{1,+}]$, cf. Fig.~\ref{fig2}(b). As intuitively explained above, the nonreciprocal statistical phenomena of the magnons in the qubit-magnon system stem from both the coherent and dissipative qubit-magnon couplings.

\section{Discussions and conclusions}\label{sec_dis}

It is worth discussing the effect of the coupling between the cavity and the waveguide on the nonreciprocal magnon blockade, which is ignored in the analysis above. When considering the cavity-waveguide coupling, the master equation in Eq.~(\ref{eq_me}) is modified as
\begin{eqnarray} \label{eq_me2}
\dot{\rho}&=&i \left[\rho, H+H_d\right]+i \left[\rho,\omega_c c^\dag c\right]
                  +\gamma_{\rm in}{\mathcal L}[\sigma_-]\,\rho \nonumber \\
          & &+\kappa_{\rm in}{\mathcal L}[b]\,\rho +\beta_{\rm in}{\mathcal L}[c]\,\rho + \tau {\mathcal L}[o'] \,\rho,~~~~~
\end{eqnarray}
where $c^\dag$ and $c$ are the creation and annihilation operators of the cavity mode with the intrinsic decay rate $\beta_{\rm in}$. Here the jump operator $o$ is replaced by $o'$ as~\cite{metelmann2015prx}
\begin{eqnarray}\label{eq_opo1}
o' = e^{i\Theta}\mu b + \nu \sigma_- + \zeta c.
\end{eqnarray}
The coefficient $\zeta$, which satisfies $\tau \zeta^{2}=\beta_{\rm ex}$, denotes the coupling of the cavity mode to the traveling photons in the waveguide. Applying the same strategy of separating the quantum jump terms, the effective non-Hermitian Hamiltonian $H_{\rm eff}$ in \eq{\ref{eq_ht1eff}} is modified to include additional terms,
\begin{eqnarray}\label{eq_heffadd}
H_{\rm add} &=& -i\Gamma_1 (\sigma_{+}c +c^{\dagger} \sigma_{-})
                 -i\Gamma_2 e^{i\Theta} (b^{\dagger}c +c^{\dagger} b),
\end{eqnarray}
with $\Gamma_1=\sqrt{\gamma_{\rm ex}\beta_{\rm ex}}$ and $\Gamma_2=\sqrt{\kappa_{\rm ex}\beta_{\rm ex}}$, which represent the cavity-qubit and cavity-magnon dissipative couplings mediated by the waveguide, respectively.

\begin{figure}[bt]
\begin{center}
\includegraphics[scale=0.6]{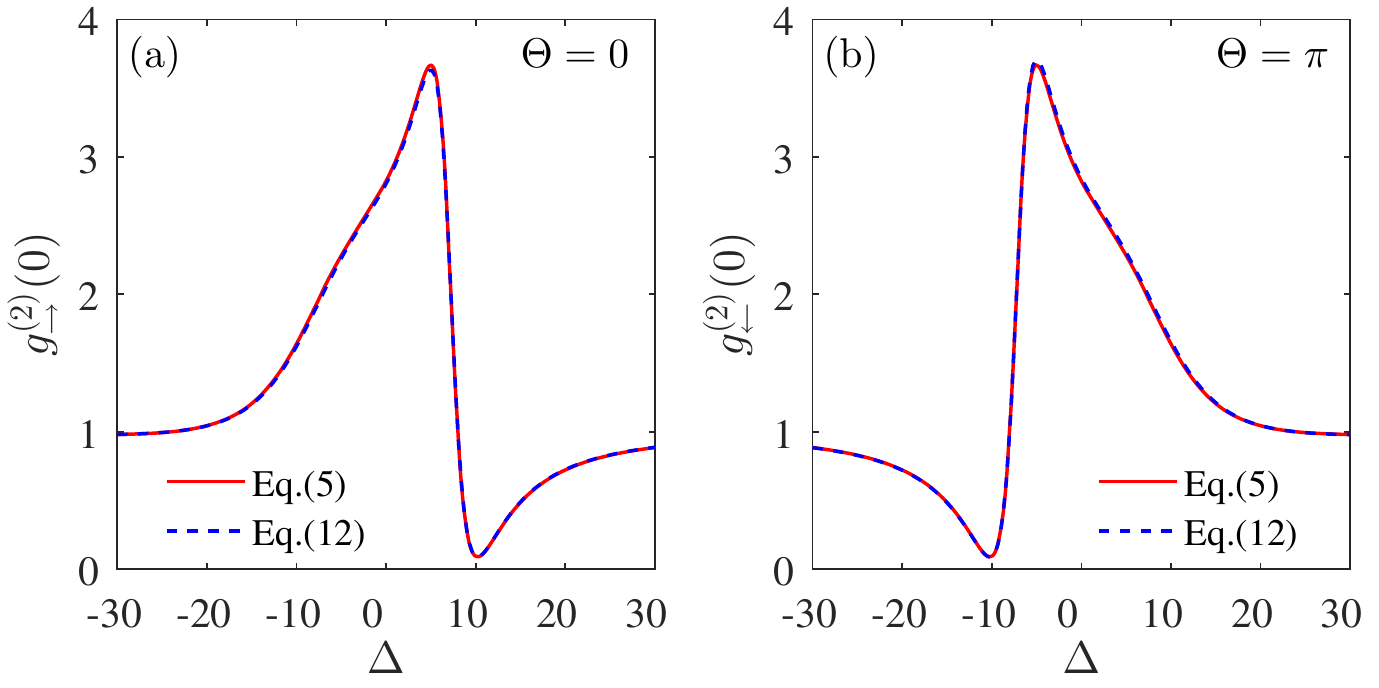}
\end{center}
\caption{Comparison of the second-order correlation function $g^{(2)}(0)$ for the cases with and without the coupling between the cavity and the waveguide, which are calculated using \eq{\ref{eq_me}} and \eq{\ref{eq_me2}}. In (a) and (b), the drivings are loaded via port 1 and port 2, respectively. The newly introduced parameters are chosen as $\beta_{\rm in}/2\pi = 1$~MHz and $\zeta=1$, while other parameters are the same as in Fig.~\ref{fig2}(c).}
\label{fig5}
\end{figure}

We then perform numerical comparisons of the second-order correlation functions $g_{\rightleftarrows}^{(2)}(0)$ using master equations in \eq{\ref{eq_me}} and \eq{\ref{eq_me2}}. In the case of either $\Theta = 0$ [c.f.~Fig.~\ref{fig5}(a)] or $\Theta = \pi$ [c.f.~Fig.~\ref{fig5}(b)], the behaviors of the red and blue curves are in good agreement with each other, even when the dissipative cavity-qubit and cavity-magnon coupling strengths are comparable to their dissipative qubit-magnon counterpart. This indicates that it is indeed reasonable to neglect the direct coupling between the cavity and the waveguide. The essential physical mechanism behind can be further described below. First, the central conductor of the microwave cavity can be structurally engineered to be spatially separated from the waveguide except for the segments around the qubit and the magnons [see Fig.~\ref{fig1}(a)]. This can leads to negligible cavity-waveguide coupling. Second, the cavity can be much detuned from both the qubit and the magnon mode, so the waveguide-induced cavity-qubit and cavity-magnon dissipative couplings can be safely neglected under the rotating-wave approximation~\cite{Scully97}. Therefore, it is sufficient to ignore the effect of the cavity-waveguide coupling on the magnon statistics.

Before concluding, we briefly assess the feasibility of the physical implementation of the present scheme. Strong coherent interaction (of strength $g/2\pi\sim 10$ MHz) between a superconducting qubit and the Kittel mode, mediated by virtual photons in the detuned microwave cavity, has been achieved when the frequency $\omega_q$ of qubit is approximately equal to the frequency $\omega_b$ of magnon mode ($\omega_q/2\pi \approx \omega_b/2\pi \approx 8.4$~GHz)~\cite{tabuchi2015s,Quirion20}. On the other hand, modest intrinsic qubit (cavity) damping rate of $\gamma_{\rm in}/2\pi \approx 1$~MHz ($\beta_{\rm in}/2\pi \approx 1$~MHz) can be chosen from Ref.~\cite{niemczyk2011,forn-diaz2019rmp}, and an intrinsic magnon damping rate of $\kappa_{\rm in}/2\pi \approx 1$~MHz can be fitted from the transmission spectra in Ref.~\cite{wang2019prl}. Moreover, an external relaxation rate of $\gamma_{\rm ex}/2\pi = 11$~MHz for the superconducting qubit, solely caused by the open transmission line, was reported by demonstrating an 94\% extinction of the light in the resonance fluorescence experiment~\cite{astafiev2010s}. In Ref.~\cite{Qian20}, the external decay rate $\kappa_{\rm ex}/2\pi = 0.33$~MHz of the magnon mode due to the waveguide was engineered, where the small external decay rate is specially designed by placing the YIG sphere $1$~mm distance away from the waveguide to obtain a high isolation ratio for the nonreciprocal microwave propagation. In fact, if the distance between the YIG sphere and waveguide is decreased to, e.g., $0.1$~mm, the external decay rate $\kappa_{\rm ex}$ of the magnon mode can be significantly increased.

To summarize, we have theoretically studied the nonreciprocal magnon blockade in a magnon-based hybrid system. Mediated by virtual photons in a microwave cavity (traveling microwaves in a waveguide), the magnon mode in a YIG sphere can be coherently (dissipatively) coupled to a superconducting qubit. By investigating the equal-time second-order correlation function of magnons and analysing the energy level structure of the hybrid quantum system, we demonstrate that the interference between the coherent and dissipative qubit-magnon couplings can yield the nonreciprocal magnon blockade. Moreover, our approach is experimentally implementable with the state-of-the-art technologies. It offers a new way to develop nonreciprocal quantum devices based on the nonreciprocal magnon blockade, and may find promising applications in chiral quantum technologies~\cite{Yang2019,Jiao2020,Dong2021,Xu2020,Xu2021}.

\vspace{8pt}
\section*{Acknowledgments}
This work was supported by the National Natural Science Foundation of China (Grants No. 11934010 and No. U1801661), the National Key Research and Development Program of China (Grant No. 2016YFA0301200), and the Zhejiang Province Program for Science and Technology (Grant No. 2020C01019).
W. X. is supported by National Natural Science Foundation of China (Grant No. 11804074). 

\appendix

\section{Coherent qubit-magnon coupling via virtual photons in the microwave cavity}\label{Appendix-A}

For a hybrid system consisting of a qubit and a YIG sphere both coupled to a cavity, the total Hamiltonian of the system can be written as
\begin{equation}\label{A1}
\begin{split}
&H_s=H_{0}+H_{I},\\
&H_{0}=\omega_q^{(0)}\sigma_{+}\sigma_{-}+\omega_b^{(0)}b^{\dag}b+\omega_{c}c^{\dag}c,\\
&H_{I}=\lambda_{q}(\sigma^{+}c+\sigma^{-}c^{\dag})+\lambda_b(b^{\dag}c+bc^{\dag}).
\end{split}
\end{equation}
In the case of $|\delta_b|\gg \lambda_b$ and $|\delta_{q}|\gg \lambda_{q}$, with $\delta_{q}=\omega_q^{(0)}-\omega_{c}$, and $\delta_b=\omega_b^{(0)}-\omega_{c}$, the effective coupling between the Kittel mode and the qubit can be obtained by applying the Fr\"{o}hlich-Nakajima transformation onto the Hamiltonian $H_s$ in Eq.~(\ref{A1})~\cite{tabuchi2015s,Quirion20}. 

We consider a unitary transformation $U= \exp(V)$, where
\begin{equation}\label{A2}
V=\frac{\lambda_{q}}{\delta_{q}}(\sigma^{-}c^{\dag}-\sigma^{+}c)+\frac{\lambda_b}{\delta_b}(bc^{\dag}-b^{\dag}c)
\end{equation}
is anti-Hermitian and satisfies $H_{I} + [H_{0}, V]  = 0$.  Up to the second order, the transformed Hamiltonian, $H=U^{\dag}H_s U$, can be approximatively written as~\cite{Frohlich50,Nakajima55}
\begin{eqnarray}\label{A3}
H&\approx &H_{0}+\frac{1}{2}[H_{I},V] \nonumber\\
&=& \omega_q\sigma_{+}\sigma_{-}+\omega_b b^{\dag}b+\lambda(b^{\dag}\sigma^{-}+b\sigma^{+})\nonumber\\
&& +\left(\omega_c-\frac{\lambda_b^{2}}{\delta_b}+\frac{\lambda_{q}^{2}}{\delta_{q}}\sigma_{z}\right)c^{\dag}c,
\end{eqnarray}
where $\omega_q = \omega^{(0)}_q +\lambda_q^2/\delta_q$, $\omega_b=\omega_b^{(0)} +{\lambda_b^2}/\delta_b$, and $\lambda={\lambda_q\lambda_b}({1}/{2\delta_q}+{1}/{2\delta_b})$. In our approach, the cavity mode is largely detuned from the Kittel mode and the superconducting qubit, so it is reasonable to assume that the cavity mode always remains in the ground state, i.e., $\langle c^\dag c\rangle \approx 0$. Under this approximation of $\langle c^\dag c\rangle \approx 0$, the effective Hamiltonian in Eq.~(\ref{A3}) can be reduced to \eq{\ref{JC}} in the main text.

\end{document}